\documentclass[letterpaper, 10 pt, conference]{ieeeconf}  

\IEEEoverridecommandlockouts                              

\usepackage{hyperref}
\usepackage{graphicx}

\usepackage{float}

\title{\LARGE \bf
Uncertainty, Vagueness, and Ambiguity in Human-Robot Interaction: Why Conceptualization Matters} 

\author{Xiaowen Sun$^{*}$, Cornelius Weber, Matthias Kerzel, Josua Spisak, and Stefan Wermter$^{}$
\thanks{*Corresponding author}
\thanks{$^{}$The Authors are with the Knowledge Technology Group, Department of Informatics, University of Hamburg, 22527 Hamburg, Germany.
        {\tt\small xiaowen.sun@uni-hamburg.de}}%
}

\begin{document}

\maketitle
\thispagestyle{empty}
\pagestyle{empty}

\begin{abstract}

Uncertainty, vagueness, and ambiguity are closely related and often confused concepts in human-robot interaction (HRI). 
In earlier studies, these concepts have been defined in contradictory ways and described using inconsistent terminology. This conceptual confusion and lack of terminological consistency undermine empirical comparability, thereby slowing the accumulation of theory.
Consequently, consistent concepts that clarify these challenges, including their definitions, distinctions, and interrelationships, are needed in HRI. To address this lack of clarity, this paper proposes a consistent conceptual foundation for the challenges of uncertainty, vagueness, and ambiguity in HRI. First, we examine the meanings of these three terms in dictionaries. We then analyze the nature of their distinctions and interrelationships within the context of HRI. We further illustrate these characteristics through examples. Finally, we demonstrate how this consistent conceptual foundation facilitates the design of novel methods and the evaluation of existing methodologies for these phenomena.

\end{abstract}

\section{Introduction}
When robots assist and cooperate with humans, human users occupy a dominant role, issuing commands to the robot through natural language.
The robot, in a subordinate role, is responsible for carrying out these tasks. To fulfill the user’s intentions, the robot must follow human instructions while also interacting autonomously with its environment to determine and execute appropriate actions.

Communication~\cite{brohan2023can, ren2023robots}, perception~\cite{shimoda2022role, qian2024affordancellm}, interaction~\cite{soni2024advancing, li2024embodied, xiao2025robot}, and execution~\cite{cui2025task, 10553231} are essential for autonomous robots to achieve this overarching goal as assistants.
Communication plays a critical role, functioning as the bridge that ensures information is properly aligned between the user and the robot.
However, environmental dynamics, the user’s linguistic limitations, or the robot’s own capabilities can hinder communication. As a result, interactions encounter problems arising from uncertainty, vagueness, and ambiguity.

A more fundamental challenge is that the problem itself is defined in contradictory ways. It is characterized by inconsistent terminology, uneven emphasis, and the absence of a coherent conceptual framework.
The literature exhibits substantial inconsistency in how key terms are defined and used.
In several studies, ambiguity and uncertainty are treated interchangeably, resulting in conceptual confusion and obscuring the precise nature of the problem. For instance, Senanayake~\cite{senanayake2024role} states that ``Uncertainty, also known as ambiguity, stems from the stochasticity of the world,'' thereby conflating the two notions.
A number of studies falsely regard the lack of clear or complete information about possible outcomes as an ambiguity problem~\cite{chisari2025robotic, liu2022perspective, morales2013ambiguity, karli2023extended, fan2024vision, davila2025llm, pramanick2022talk}.
In contrast, other studies conceptualize similar phenomena primarily under the terms of uncertainty~\cite{trick2019multimodal, hu2022active, kwon2020humans, scherf2024you}.
Meanwhile, the notion of vagueness is often overlooked or implicitly subsumed under either ambiguity or uncertainty.
To the best of our knowledge, the collaborative behavior-based approach of Wang et al.~\cite{wang2006interface} is the only study that explicitly considers ambiguity, uncertainty, and vagueness distinctly in the context of service robots. However, their study focuses solely on natural language sentence parsing, thereby overlooking the dynamics and complexity of the environment.
This chain reaction affects the scholarly community: a false definition leads to conceptual confusion, which produces inconsistent studies, undermines empirical comparability, and ultimately slows theoretical accumulation.

\begin{figure}
    \centering
    \includegraphics[width=1\columnwidth]{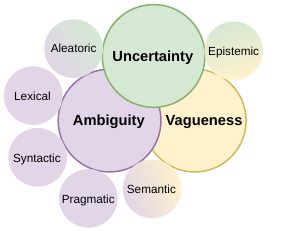}
    \caption{The relationship of uncertainty, vagueness, and ambiguity in the interdisciplinary research of linguistics and HRI. 
    \textit{Uncertainty} refers to states of incomplete knowledge. \textit{Vagueness} pertains to concepts with indeterminate boundaries. \textit{Ambiguity} arises when a linguistic or symbolic expression maps to two or more distinct interpretations. These three phenomena can occur independently or in combination, as illustrated by the smaller circles.
    }
    \label{fig: problems}
    \vspace{-0.3cm}
\end{figure}

Accordingly, a more systematic examination of uncertainty, vagueness, and ambiguity is required within the field of HRI. \textbf{In this paper, we establish a consistent conceptual foundation (see Fig.~\ref{fig: problems}) about these challenges}. We will outline this foundation in the remainder of the paper.
\vspace{0.3cm}

\begin{table*}[t]
    \centering
     \resizebox{1\textwidth}{!}{
    \begin{tabular}{p{1.8cm}| p{4cm} | p{8cm}} %
      \multicolumn{1}{c|}{\textbf{Type}} & 
      \multicolumn{1}{c|}{\textbf{Instruction}} & 
      \multicolumn{1}{c}{\textbf{Explanation}} \\
      \hline
      Uncertainty& \texttt{Pick up the toy car from under the table.}& Due to the toy car not being in the robot's view, the robot is \textit{uncertain} whether it is under the table or not. \\
      \hline
       Ambiguity  & \texttt{Pick up the apple on the table.} & There are multiple apples. The instruction is \textit{ambiguous} because it could refer to more than one object.\\
       \hline
       Vagueness   & \texttt{Clean this table thoroughly.} & `Thoroughly' is not precisely defined; boundaries of the task are unclear, leading to \textit{vagueness} in this command.\\
       \hline
       Uncertainty and Ambiguity & \texttt{Deliver this glass to another room.} & There are many rooms; `another' is \textit{ambiguous}. From the robot's current location to another room, the environment is dynamic, with people and obstacles moving unpredictably. The robot cannot predict exactly where obstacles will be; there is \textit{uncertainty}.\\
       \hline
      Vagueness and Ambiguity & \texttt{Move closer to the table.}  & There are multiple tables in the room, and `closer' is not quantified. This command combines \textit{ambiguity} (which table?) and \textit{vagueness} (how close?).\\
        \hline
      Uncertainty and Vagueness  & \texttt{Can you clean this messy area later?} & The words, `this messy area' and `later' lack clear boundaries. In a very messy scene, the robot may lack the capability to clean it. Therefore, this command contains \textit{vagueness} and \textit{uncertainty}. \\
      \hline
      Uncertainty, Vagueness, and Ambiguity & \texttt{Bring me the small box from the other room later.} & The instruction contains \textit{ambiguity} (`the small box,' `the other room'), \textit{vagueness} (`later'), and \textit{uncertainty} due to the dynamic and unpredictable nature of the environment.\\
      \hline
      & \textbf{$\cdots$ }&\\
    \end{tabular}}
    \caption{
    This table illustrates UVA-phenomena, distinct yet partially overlapping concepts in everyday life (see Fig.~\ref{fig: scenario}). Instructions are what a user tells the robot, while explanations provide the background and associated problems.
    }
    \label{tab: examples}
\vspace{-0.4cm}
\end{table*}

\section{Conceptual Foundation: Uncertainty, Vagueness, and Ambiguity}
\subsection{Dictionary Definitions}

Our guiding principle is that the concepts of uncertainty, vagueness, and ambiguity in HRI should be grounded, to the greatest extent possible, in the meanings of the terms themselves.
Therefore, we first review the definitions of these three words provided by the  Oxford\footnote{\url{https://www.oxfordlearnersdictionaries.com/}} and Cambridge\footnote{\url{https://dictionary.cambridge.org/dictionary/english/}} dictionaries.
\textit{Uncertainty} refers to the state of being uncertain, or a situation that causes one to feel unsure or doubtful.
\textit{Vagueness} describes something that is unclear in a person’s mind, lacking sufficient detail or information, indicating a lack of clear thought or attention, or lacking a well-defined shape or form.
\textit{Ambiguity} is defined as the state of having more than one possible meaning, or a word or statement that can be interpreted in multiple ways. It can also refer to something that is difficult to understand or explain due to its complexity or the involvement of many different aspects.

\subsection{Interpretation in HRI}

The concepts of \textit{Uncertainty}, \textit{Vagueness}\footnote{Our focus is on the linguistic and HRI aspects of vagueness, as ``fuzziness'' and ``vagueness'' are often confused in the academic literature.}, and \textit{Ambiguity} (UVA-phenomena) manifest in different communicative and decision-making processes in HRI. 
UVA-phenomena denote distinct but partially overlapping concepts, which means that these challenges can occur independently or in combination~\cite{kennedy2011ambiguity, tuggy1993ambiguity}. 

\begin{figure}[H]
    \centering
\includegraphics[width=1\columnwidth]{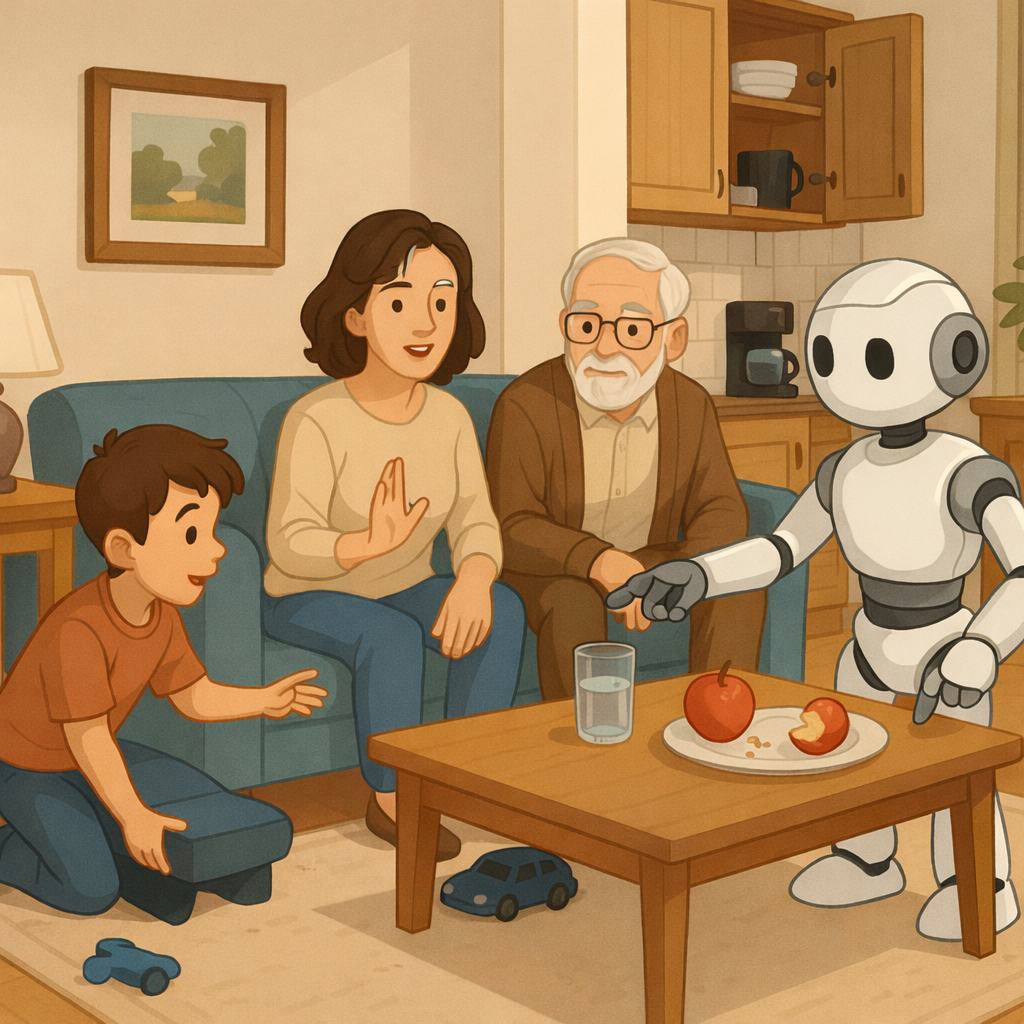}
    \caption{An everyday life scenario involving a humanoid intelligent robot.
    Source: AI-generated via ChatGPT.
    }
    \label{fig: scenario}
    \vspace{-0.3cm}
\end{figure}

Their tight relationships are depicted in Fig.~\ref{fig: problems} by large circles, and their ``distinct but partially overlapping'' nature is illustrated in smaller circles.
Concrete examples in Tab.~\ref{tab: examples} further demonstrate this nature.

Our objective is to develop an intelligent agent that assists humans with everyday activities, which requires the agent to be capable of handling UVA-phenomena. To this end, we adopt a modeling perspective centered on the agent. Drawing on the probabilistic theory of machine learning, we categorize \textit{uncertainty} into two types: epistemic and aleatoric uncertainty~\cite{pml1Book, abdar2022uncertainty}.
Epistemic uncertainty refers to whether the agent possesses sufficient cognitive capabilities and has been provided with adequate knowledge.
Aleatoric uncertainty arises from the inherent stochasticity of a dynamic environment or event.

\textit{Vagueness} can be divided into epistemic vagueness and semantic vagueness.
Epistemic vagueness arises from the agent’s limited knowledge or incomplete information about the environment. In this case, it aligns with epistemic uncertainty.
Semantic vagueness originates from the inherent imprecision of the concepts or language used to describe the task or environment.

\textit{Ambiguity} can arise either from the language itself or from its mapping to the environment.
Language can be analyzed at the lexical, syntactic, contextual, and logical levels. 
Accordingly, ambiguity can be classified as lexical, syntactic, pragmatic (stemming from context), or semantic (stemming from logic)~\cite{kennedy2011ambiguity, vanRooij2011}. 
In HRI, natural language instructions often involve multiple interpretations of words, phrases, or sentence structures. Even when grammatically correct, such instructions can yield different interpretations depending on the environment.
Moreover, users’ linguistic capabilities, for instance, children or elderly individuals who may produce ungrammatical instructions, can introduce additional ambiguity.
Additionally, environmental dynamics can cause aleatoric ambiguity. For example, a robot may fail to accurately recognize a user’s speech due to external random noise that may appear only temporarily.
We align this phenomenon with aleatoric uncertainty.

Each of the UVA-phenomena arises from a specific source or cause. In the case of \textit{uncertainty}, the primary source is the agent's lack of information, whether due to limitations in internal cognition or the absence of external data.
The underlying cause of \textit{vagueness} is the imprecise boundaries of concepts, which may result from the user's subjective cognition or from objective characteristics.
In contrast, the root cause of \textit{ambiguity} lies in the user's linguistic instructions, particularly when they refer to the environment, leading to multiple possible interpretations~\cite{manning1999foundations}.

\subsection{Application in HRI}

A detailed conceptualization of UVA-phenomena facilitates the development of solutions in HRI, as effective solutions must be grounded in the underlying nature of the problem.
UVA-phenomena characterize this nature through aleatoric, epistemic, semantic, lexical, syntactic, and pragmatic factors that models should consider.
Models addressing aleatoric aspects should incorporate mechanisms for handling uncertainty and ambiguity.
Models focusing on epistemic aspects should address uncertainty and vagueness, whereas models targeting semantic aspects should consider ambiguity and vagueness.
Modeling lexical, syntactic, and pragmatic factors is also essential for dealing with ambiguity.

For example, consider the uncertainty problem illustrated by the instruction \texttt{Pick up the toy car from under the table} in the scenario shown in Fig.~\ref{fig: scenario}. The robot cannot see the toy car at its current location. This represents an epistemic uncertainty problem caused by insufficient information.
The next action should therefore be to move in order to obtain additional visual information. 
It also demonstrates that actions can be derived from clarifying UVA-phenomena.
In contrast, consider the instruction \texttt{Deliver this glass to another room} in the same scenario. This instruction involves both uncertainty and ambiguity. The phrase `another room' introduces semantic ambiguity because multiple rooms are possible destinations. Furthermore, moving from the current location to another room involves aleatoric uncertainty due to the dynamic environment.

UVA-phenomena also serve as a principle for systematically analyzing existing methodologies.
For example, the KnowNo framework~\cite{ren2023robots} integrates the generation of multiple candidate actions, conformal prediction, and downstream decision-making to jointly handle ambiguity and uncertainty. However, it overlooks the phenomenon of vagueness.
The HYNA architecture~\cite{sun2022learning} utilizes an LSTM network to track the multimodal dialog state. It focuses on handling ambiguity in users’ linguistic instructions and environmental contexts; however, it does not specifically address vagueness. The OSSA framework~\cite{sun2024state} leverages pretrained foundation models for robotic task planning, focusing on domestic robots that autonomously complete both short- and long-horizon tasks while considering user preferences. It can handle both ambiguous and vague situations.
The robot executes the instruction \texttt{clear the table} in the scenario shown in Fig.~\ref{fig: scenario}. OSSA first distinguishes between an intact apple and a cut apple and then generates different plans accordingly: the intact apple is stored in the cabinet, whereas additional information is requested for the cut apple.

We demonstrate how UVA-phenomena can be used to evaluate existing studies. However, a more comprehensive review is needed to provide an overall summary of the development of UVA-phenomena in HRI.

\section{Conclusion and Future Work}

In this paper, we establish a conceptual foundation for uncertainty, vagueness, and ambiguity (UVA-phenomena) in HRI. Grounded in dictionary definitions, we interpret and demonstrate the nature of UVA-phenomena in HRI, which exhibit distinct yet partially overlapping characteristics. We further show how this conceptual foundation provides a structured framework for systematically addressing UVA-phenomena in the development of future HRI models and for evaluating existing methodologies.

By explicitly modeling and clearly communicating uncertainty, vagueness, and ambiguity, researchers can strengthen the foundations of trustworthy artificial intelligence (AI), as these factors directly affect transparency, reliability, safety, and the calibration of user trust. Establishing a shared conceptual understanding of UVA-phenomena promotes methodological consistency across studies, enhances empirical comparability, and facilitates cumulative theoretical development within HRI. Collectively, these contributions support advancing HRI toward greater robustness, reliability, and intelligence.

From a broader AI perspective, UVA-phenomena extend beyond HRI to agentic AI, robotics, and signal processing. It may further inform research in other domains, such as cognitive science.


\section*{ACKNOWLEDGMENT}

The authors gratefully acknowledge support from the China Scholarship Council (CSC) and the German Research Foundation DFG under project CML (TRR 169).



\bibliographystyle{IEEEtran}
\bibliography{reference}

\end{document}